\documentclass[pre,aps,floats,superscriptaddress,floatfix,twocolumn,longbibliography]{revtex4-1}
\usepackage{amssymb,amsmath}
\usepackage{amsmath,amssymb}
\usepackage{enumitem,kantlipsum}
\usepackage{lipsum}

\usepackage{gensymb}
\usepackage{graphicx}
\usepackage{psfrag}
\usepackage{color}
\usepackage{dcolumn}
\usepackage{bm}
\usepackage[normalem]{ulem}
\usepackage[absolute]{textpos}
\usepackage{hyperref}
\hypersetup{
  colorlinks=true,
  citecolor=magenta,
  linkcolor=cyan,
}

\begin{document}
\title{
Rough or crumpled: Strong coupling phases of a generalized Kardar-Parisi-Zhang surface
}
\author{Debayan Jana}\email{debayanjana96@gmail.com}
\affiliation{Theory Division, Saha Institute of Nuclear Physics, A CI of Homi Bhabha National Institute,  1/AF Bidhannagar, Calcutta 700064, West Bengal, India}
\author{Abhik Basu}\email{abhik.123@gmail.com, abhik.basu@saha.ac.in}
\affiliation{Theory Division, Saha Institute of Nuclear Physics, A CI of Homi Bhabha National Institute,  1/AF Bidhannagar, Calcutta 700064, West Bengal, India}

\begin{abstract}

We study a generalized Kardar-Parisi-Zhang (KPZ) equation [Jana {\em et al}., Phys. Rev. E {\bf 109}, L032104 (2024)] that sets the paradigm for universality in  roughening of growing
nonequilibrium surfaces without any conservation laws but with competing local and nonlocal nonlinear effects. This equation in two dimensions exhibits two distinct strong coupling regimes: a rough phase  and a crumpled phase, in addition to a weak coupling phase. The conformation fluctuations of such a rough surface are given by nonuniversal scaling exponents, with orientational long-range order and positional short-range order, whereas the crumpled phase has positional and orientational short-range order. 
\end{abstract}

\maketitle


{ 

The Kardar-Parisi-Zhang (KPZ) equation~\cite{kpz,kpz1,stanley}, originally proposed as a model for growing surfaces without overhangs, provides a paradigm  for nonequilibrium phase transitions. It undergoes a roughening transition between a smooth phase, statistically identical to a surface described by the linear Edwards-Wilkinson (EW) equation~\cite{ew}, and a perturbatively inaccessible strong coupling phase~\cite{stanley}, in dimension $d>2$. Perturbative dynamic renormalization group (RG)~\cite{stanley,tauber} analysis has been successful in describing this roughening transition in  an $\epsilon$ expansion, where $d=2+\epsilon$, with two dimensions (2D) being the lower critical dimension of the model. In $d\leq 2$, there exists only the strong coupling phase. A KPZ surface described by a single-valued height field $h({\bf x},t)$ measured with respect to an arbitrary base plane in the Monge gauge~\cite{mong} displays universal scaling in the long wavelength limit, characterized by $z$ and $\chi$,  the dynamic and roughness exponents~\cite{stanley}. These are formally defined through the correlation function
 \begin{eqnarray}
C(r,t)\equiv \langle [h({\bf x},t)-h({\bf 0},0)]^2\rangle \sim r^{2\chi}\phi ({r^z}/{t})
\end{eqnarray} 
in the scaling limit,
where $r=|{\bf x}|$ and $\phi$ is a dimensionless scaling function of its argument~\cite{stanley}. 
{In the smooth phase, $z=2$ and $\chi=1-d/2$~\cite{stanley}, and at the roughening transition at $d=2+\epsilon,\,\epsilon>0$, $z=2$ and $\chi=0$~\cite{stanley,tauber,cole}}, which are accessible within perturbative RG calculations. In contrast, the strong coupling rough phase cannot be accessed within standard RG calculations~\cite{tauber}. Failure of RG to capture the rough phase has led to the development of alternative techniques. Notable are the self-consistent mode-coupling theories (MCT)~\cite{mct_cates,jkb-mct}, which 
have predicted a variety of results on scaling within specific different calculational approaches. For example  Ref.~\cite{mct_frey} explored scaling at $d=1$, yielding  $z=3/2$, a value known {\em exactly} due to the Galilean invariance and a fluctuation-dissipation theorem~\cite{stanley}. Furthermore, Ref.~\cite{mct_moore} showed that $z$ rises from $z=3/2$ at $d=1$ to $2$ around $d_U=3.6$, where $d_U$ is the upper critical dimension of the KPZ equation defined by the dimension in which $z$ touches the value 2 from a smaller numerical value at lower $d<d_U$. In contrast, Ref.~\cite{mct_tu} suggests $z<2$ at any finite dimension indicating $d_U=\infty$. Furthermore, confirming earlier functional RG results of Ref.~\cite{healy-PRA}, Refs.~\cite{jkb-mct,jkb-book,bmhd1,bmhd2} used MCT to
arrive, quite separately, at the same dimension-dependent exponents $\chi=(4-d)/6, z=(8+d)/6$,
reasonably close to the more recent nonperturbative RG predictions~\cite{nonpert3}. This common
functional RG and MCT finding suggests explicitly that $d_U=4$, supported independently by Ref.~\cite{foged}, using a third distinct technique.
See also Ref.~\cite{mct_colaiori}, which gives $d_U=4$ and the dynamic exponent $z$ in different dimensions. Furthermore, Ref.~\cite{canet} illustrated the MCT solution for the KPZ equation at $d\geq 2$ that continues to $d_U=4$. More recent extensive numerical studies, however, appear to differ from the MCT predictions on the scaling exponents~\cite{num1,num2,num3} and suggest $d_U>4$~\cite{num11,num22,num33,num44}. See also Ref.~\cite{halp-2025} for a very recent study on the scaling in the KPZ equation at $d>2$ by combining a variety of techniques. Furthermore, the results in Ref.~\cite{num55} tend to indicate the absence of a finite $d_U$.

The dynamics of a KPZ surface
depends {\em locally} on surface fluctuations and hence cannot model nonequilibrium surfaces with nonlocal dynamics. Nonlocal effects, however, can be important in wide-ranging systems, including biological growth processes~\cite{bio}, fast nonlocal transport~\cite{transport}, and nonlocal stabilization of surfaces~\cite{krug-nonloc,nicoli}. Recently, a generalized KPZ equation (hereafter G-KPZ equation) with nonlocal effects as a conceptual model with {\em competing} local and nonlocal nonlinear effects has been proposed~\cite{active_kpz}:
\begin{align}
 \frac{\partial h}{\partial t} &= \nu \nabla^2 h+\frac{\lambda}{2} (\boldsymbol\nabla h)^2 \notag \\
 +&\quad  \lambda_1 \int d^dr' Q_{ij}({\bf r-r'})(\nabla_ih ({\bf r'},t)\nabla_jh({\bf r'},t)) + \eta. \label{ch-kpz}
\end{align}
Here, $Q_{ij}({\bf r})$ denotes the longitudinal projection operator, which in Fourier space is expressed as $Q_{ij}({\bf k})=k_ik_j/k^2$, with ${\bf k}$ being a Fourier wave vector. Thus $Q_{ij}({\bf r})$, which is the inverse Fourier transform of $Q_{ij}({\bf k})$, is nonlocal in space. Physically, the term $\lambda_1 \int d^dr' Q_{ij}({\bf r-r'})(\nabla_ih ({\bf r'},t) \nabla_jh ({\bf r'},t))$ contributes to the surface velocity normal to the base plane, $v_p=\partial h/\partial t$, and  is {\em nonlocal} in ${\boldsymbol\nabla} h$. Furthermore, $\eta({\bf x},t)$ is a zero-mean Gaussian-distributed white noise with a variance
$\langle \eta({\bf x},t)\eta({\bf 0},0) \rangle = 2D \delta^d({\bf x}) \delta(t). $ 
If $\lambda_1=0$, Eq.~\eqref{ch-kpz} reduces to the usual KPZ equation. Equation~\eqref{ch-kpz} serves as a minimal nonequilibrium model to study competition between local and nonlocal nonlinearities. For instance, as a consequence of these competitions, the G-KPZ equation (\ref{ch-kpz}) shows macroscopic properties dramatically different from the KPZ equation. It can have a stable, weak coupling sub- or superlogarithmically  rough surface in 2D, characterized by {\em nonuniversal} scaling exponents. More strikingly and unlike the 2D KPZ equation, it also admits a roughening transition even in 2D, again characterized by nonuniversal exponents. Furthermore, at $d>2$ it shows a roughening transition {\em different from that in the KPZ equation}; see Ref.~\cite{active_kpz},  which principally focused on the weak coupling phase behaviors.  Equation~\eqref{ch-kpz} admits a pseudo-Galilean invariance under the transformation $x_i'=x_i-(\lambda+2\lambda_1)c_it$ and $t'=t$, together with $h'({\bf x}',t')=h({\bf x},t)+\textbf{c}\cdot\textbf{x}$, which ensures that the combination $\lambda+2\lambda_1$ does not renormalize; see Supplemental Material~\cite{supple}, see also Ref.~\cite{active_kpz}. This further means $\chi+z=2$ {\em exactly}, as long as nonlinearities remain relevant (in the scaling sense).

In this Letter we focus on the strong coupling phases of (\ref{ch-kpz}). We show that these strong coupling phases can be of {\em two distinct} kinds: (i) a rough phase with positional short-range order (SRO) but orientational long-range order (LRO), or (ii) crumpled with both positional and orientational SRO. We extract the scaling exponents of the rough phase in $d\geq 2$. The variance $\Delta \equiv \langle [h({\bf x},t)-\overline h(t)]^2\rangle \sim L^{2\chi(\gamma)}$, $0<\chi(\gamma)<1$, where $L$ is the linear size of the surface and $\overline h(t)$ is the average height at time $t$. Furthermore, the timescale of relaxation $\tau(L)\sim L^{z(\gamma)}$. Both $\chi$ and $z$ are parametrized continuously by $\gamma\equiv \lambda_1/\lambda$, reflecting model parameter-dependent, nonuniversal scaling. In the rough (crumpled) phase, $0<\chi<1\,(\chi>1)$.

\begin{figure*}[htb]
 \includegraphics[width=\textwidth]{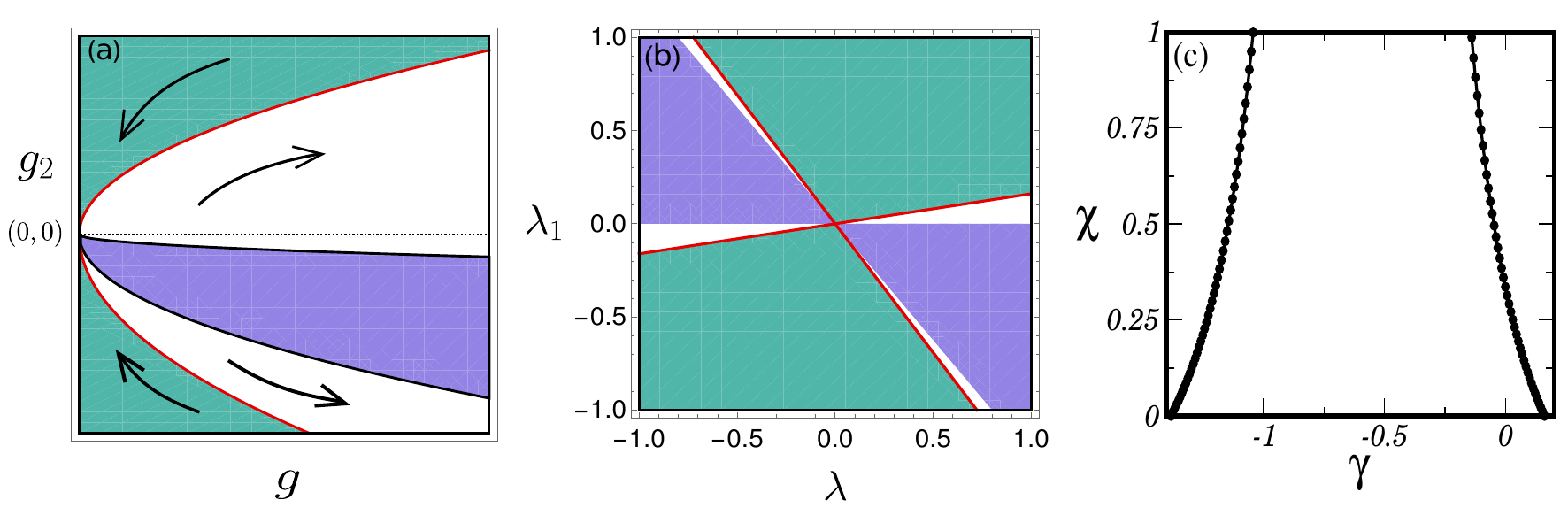}
 \caption{(a) Phases and the RG flow lines in the $g$-$g_2$ plane at 2D. Stable and unstable regions are divided by the separatrices (red lines). Stable (unstable) regions are indicated by flow lines moving towards (away from) the origin. The arrows indicate the flow directions, delineating the stable and unstable regions. In this flow diagram (0,0) is the {\em only} fixed point, which is unstable along the $g$ direction but stable along the $g_2$ direction. (b) Phases in the $\lambda$-$\lambda_1$ plane. The green regions are the weak coupling logarithmically rough phase, the white regions are the algebraically rough phase, and the purple regions are the crumpled phase. In (a) and (b) the solid red line represents the locus of unstable fixed points where the roughening transition occurs.  (c) Variation of $\chi$ in 2D in the phase space region where it is positive, with $\gamma$ as obtained from MCT. The central region where $\chi>1$ corresponds to a crumpled surface. See text.
 }\label{flow-2d}
\end{figure*}

We now derive the above results. Nonlinear terms preclude any exact analysis. We first use a Wilson dynamic RG framework~\cite{kpz,stanley,forster,tauber} and build upon the results discussed in Ref.~\cite{active_kpz}. Dimensional analysis allows us to define two effective dimensionless coupling constants:
\begin{equation}
 g\equiv \frac{\lambda^2 D}{\nu^3}k_d,\,g_1\equiv \frac{\lambda_1^2 D}{\nu^3}k_d,\,g_2\equiv \text{sgn}(\gamma)\sqrt{g_1}.\label{coup}
\end{equation}
{ Here $k_d=\frac{S_d}{(2\pi)^d}$, and $S_d$ is the surface area of a $d$-dimensional unit sphere.}
Without any loss of generality, we are considering $\lambda>0$.  We notice that from (\ref{coup}) above that $g_2$ can be both positive and negative, depending upon the sign of $\gamma$.
There are no one-loop corrections  to $\lambda,\lambda_1$. The one-loop corrections to $\nu$ and $D$ can be obtained following Refs.~\cite{stanley,forster,active_kpz}, which 
give the one-loop RG flow equations for $g$ and { $g_2$}:{
\begin{eqnarray}
 \frac{dg}{dl}&=&g[2-d-g\tilde A(\gamma)],\label{flow-g}\\
  \frac{dg_2}{dl}&=&\frac{g_2}{2}[2-d-g\tilde A(\gamma)],\label{flow-g*}
\end{eqnarray}
where $\tilde A(\gamma)=\frac{9}{8}\gamma^2+\frac{11}{8}\gamma-\frac{1}{4}$.} Considering 2D, { the physically relevant dimension}, we focus on the RG flow lines in the $g$-$g_2$ plane, with (0,0) as the only RG fixed point. Its stability properties are intriguing. It is stable (i.e., attractive) along the $g_2$ axis but unstable (i.e.,
repulsive) along the $g$ axis. This indicates the existence of a separatrix, an invariant manifold
under RG in the $g$-$g_2$ plane, that separates the stable phase from instability. While (0,0) is the {\em only} fixed point of both (\ref{flow-g}) and (\ref{flow-g*}), it may be stable or unstable. Stability requires $dg/dl<0, dg_2/dl<0$, so that {\em both} $g(l),\,g_2(l)$ flow to zero in the long wavelength limit. Otherwise, when $dg/dl>0, dg_2/dl>0$, both $g(l),\,g_2(l)$ flow {\em away} from (0,0), giving instability. These two behaviors are demarcated by the separatrices $\tilde A(\gamma)=0$,
giving 
\begin{equation}
g_2=(0.161)\sqrt g,\; g_2=-1.383\sqrt g \label{sep-eq1}
\end{equation}
for $\lambda_1>0$ and $\lambda_1<0$, respectively, as the separatrices, in a $g$-$g_2$ plane passing through (0,0). The RG flow diagram in 2D is shown in Fig.~\ref{flow-2d}(a). The red lines in Fig.~\ref{flow-2d}(a) are the separatrices given in (\ref{sep-eq1}). The RG flow lines in the green regions demarcated by the separatrices flow to the origin, giving the ``weak coupling phase.'' Although (0,0) is the stable fixed point in this region, the flow to (0,0) is so slow that $\nu$ and $D$ are infinitely renormalized, giving super- or sublogarithmically rough phases, as explored in Ref.~\cite{active_kpz}. In the remaining regions, both $g$ and $g_2$ grow monotonically with the ``RG time'' $l$. As soon as $g(l),\,g_2(l)\sim {\cal O}(1)$, which happens for a surface of finite size, the RG flow equations are outside the validity of the perturbative expansions. This is the perturbatively inaccessible ``strong coupling phase.''

We first argue that the region confined between the two separatrices [red lines in Fig.~\ref{flow-2d}(a)] actually consists of two subregions with distinct scaling properties (instead of just one strong coupling phase, as originally speculated in Ref.~\cite{active_kpz}). To investigate that, first consider the bare perturbation theory for $\nu$ in 2D, which is same as the RG calculations for $\nu$ except that now we extend the integrals over wave vectors down to an infrared cutoff $q_\text{min}=2\pi/L$. Setting $d=2$, we get for the effective $\nu_e$,
\begin{eqnarray}
 &&\nu_e \approx 
 \nu_0 +g\nu_0\Bigl(\frac{1}{2}\gamma^2+\frac{5}{8}\gamma\Bigl)\ln (L/a). \label{nu-e}
\end{eqnarray}
We thus find that for $\Bigl(\frac{1}{2}\gamma^2+\frac{5}{8}\gamma\Bigl)<0$ or $\Bigl(\frac{(g_2)^2}{2g}+\frac{5g_2}{8\sqrt g}\Bigl)<0$, {$-1.25<\gamma<0$}, $\nu_e<\nu_0$. For a crumpled surface, $\langle \delta {\bf n}^2\rangle$ diverges for sufficiently large $L/a$, where $\delta {\bf n}$ is deviation of the local normal from a fully flat state. In the Monge gauge~\cite{mong}, to the lowest order in height fluctuations $\delta n\sim {\boldsymbol \nabla }h$. For sufficiently large $L/a$, $\nu_e<0$ gives divergence of $\langle ({\boldsymbol\nabla} h)^2\rangle$, implying surface crumpling with positional and orientational SRO. 
In higher dimensions $d>2$, the one-loop integrals are finite and hence the lower limit can be extended to 0, corresponding to the thermodynamic limit. Then for a range of $\gamma$ given by {$-1.25<\gamma<0$} and low enough bare $\nu_0$, $\nu_e$ can turn negative, once again indicating crumpling. We will revisit the region with crumpling below, shown in purple in Fig.~\ref{flow-2d}(a), by using mode coupling methods and come to a similar conclusion, giving us confidence about the existence of a crumpled phase. See Fig.~\ref{flow-2d}(b) for a schematic phase diagram in the $\lambda$-$\lambda_1$ plane.

To study the unstable regions in the RG flow diagrams, we 
now set up a one-loop MCT calculation~\cite{jkb-mct,bmhd2} and extract the scaling behavior by using it; see also~\cite{akc-ab-jkb,abmhd,abjkb} for applications of MCT in slightly different but related contexts. 
The response $G({\bf k},\omega)$ and correlation $C({\bf k},\omega)$ functions have the following scaling form: 
\begin{align}
&G^{-1}({\bf k},\omega)=k^{z}g\biggl(\frac{\omega}{k^z}\biggr)=-i\omega+\Sigma({\bf k},\omega),\label{resp1}\\
&C({\bf k},\omega)=k^{-d-2\chi-z}f\biggl(\frac{\omega}{k^z}\biggr),
\end{align}
where the self-energy $\Sigma ({\bf k},\omega=0)=\Gamma k^z$ in the zero frequency limit. Similarly for the correlator in the Lorentzian approximation, we write
\begin{align}
C({\bf k},\omega)=\frac{2D\Gamma k^{-2\chi-d+z}}{\omega^2+\Gamma^2k^{2z}}.\label{correl_1}
\end{align}
Our argument is that the universal amplitude ratio $\Gamma^2/(D\lambda^2)$ can be written down both from the one-loop diagrammatics of $G^{-1}({\bf k},\omega)$ and $C({\bf k},\omega)$. We use $\chi+z=2$, which is an {\em exact} result in the strong coupling phases  (also consistent with the absence of vertex corrections at the one-loop order), further assuming that $G^{-1}$ and $C$ are dominated by their respective one-loop contributions,which would hold if $z<2$ and $\chi>0$. Combining contributions from all the one-loop diagrams for $G({\bf k},\omega=0)$, we obtain~\cite{supple}
\begin{align}
\frac{\Gamma^2}{\lambda^2D}&=\frac{k_d}{\chi\, d}\biggl[\frac{\chi}{2}+\gamma\biggl(\frac{4d+2d^2+2\chi d+10\chi-6}{2(d+2)}\biggl)\nonumber\\&+\gamma^2\biggl(\frac{6\chi+4d-4}{d+2}\biggl)\biggl].\label{relation_1}
\end{align}
From the one-loop contributions to $C({\bf k},\omega=0)$, we find
\begin{align}
&\frac{\Gamma^2}{\lambda^2D}=\frac{k_d\biggl[d(d+2)+4\gamma(d+2)+12\gamma^2 \biggl]}{4d(d+2)(d+3\chi-2)}.\label{relation_2}
\end{align} 
Now comparing the right-hand side of Eqs.~(\ref{relation_1}) and~(\ref{relation_2}), we obtain a quadratic equation of  $\chi$: 
\begin{align}
A\chi^2+B\chi+C=0.\label{chi-eq}
\end{align}
 Here,
$A=36\gamma^2+(6d+30)\gamma+(3d+6),\,B=(36d-54)\gamma^2+(8d^2+16d-42)\gamma+\biggl(\frac{d^2}{2}-4-d\biggl),\,C=(d-2)\gamma\bigl[\gamma(8d-8)+(2d^2+4d-6)\bigl].$
Solving (\ref{chi-eq}),
\begin{align}
\chi=\frac{-B\pm|\sqrt{B^2-4AC}|}{2A}=\chi_\pm.\label{chi-val}
\end{align}
We take $\chi=\chi_+$, for with $\gamma=0$ only  $\chi_+$ reduces to  $\chi=1/2$ at $d=1$, the known exact value; incidentally, $\chi=\chi_+$ reduces to $\chi=1/3$ in 2D and $d_U=4$ obtained in an analogous MCT for the pure KPZ equation~\cite{jkb-mct} for $\gamma=0$. We now calculate $\chi$ in 2D. With  $d=2$ in   $A$, $B$, and $C$ defined above, we get $C=0$ and $\chi=\frac{-B+|B|}{2A}$, where $B=18\gamma^2+22\gamma-4$ and $A=36\gamma^2+42\gamma+12$. Validity of our MCT requires $\chi>0$. This means either $B<0$ and $A>0$, or $B>0$ and $A<0$. Notice that $B=0$ gives the separatrix 
(\ref{sep-eq1}) with $B<0$ corresponding to the unstable region in Fig.~\ref{flow-2d}(a). We focus on the region $B<0$, which holds for $-1.383<\gamma<0.161$. Within this region,
\begin{align}
\chi=\frac{-B}{A}=\frac{-18\gamma^2-22\gamma+4}{36\gamma^2+42\gamma+12}.\label{chi-val-2d}
\end{align}} 


On the separatrices (\ref{sep-eq1}), $B=0$ and $A=22.77$ { corresponding to $\gamma=\gamma_{c1}=0.161$} and $A=19.69$, { corresponding to $\gamma=\gamma_{c2}=-1.383$}; $A$ changes from these boundary values within the unstable region between the separatrices. In particular,
as $A$, depending upon $\gamma$, decreases,  $\chi$ grows, eventually exceeding unity.  This gives   $\langle ({\boldsymbol\nabla}h)^2\rangle$ diverging with $L$, giving a crumpled surface. [We make a technical point here that as $\chi$ exceeds unity, higher order nonlinear terms, e.g., $({\boldsymbol \nabla} h)^{2n},\,n>1$, not included in (\ref{ch-kpz}) should
be relevant (in the scaling/RG sense). Thus Eq.~\eqref{ch-kpz}, though it predicts the onset of the crumpled phase, cannot be used to study the crumpled phase.] Setting $\chi=1$ in \eqref{chi-val-2d}, we get $\gamma=-1.043,-0.142$. { Thus in the region $(-1.043<\gamma<-0.142)$ MCT predicts a crumpled phase, which is the purple region in Fig.~\ref{flow-2d}(a) and is slightly different from the region defined by the condition $\nu_e=0$ [see Eq.~(\ref{nu-e}) above].} 
Due to the breakdown of our theory, we cannot make any definitive conclusion about any scaling properties in the crumpled region. 

\begin{figure}[b]
 \includegraphics[width=0.32\textwidth]{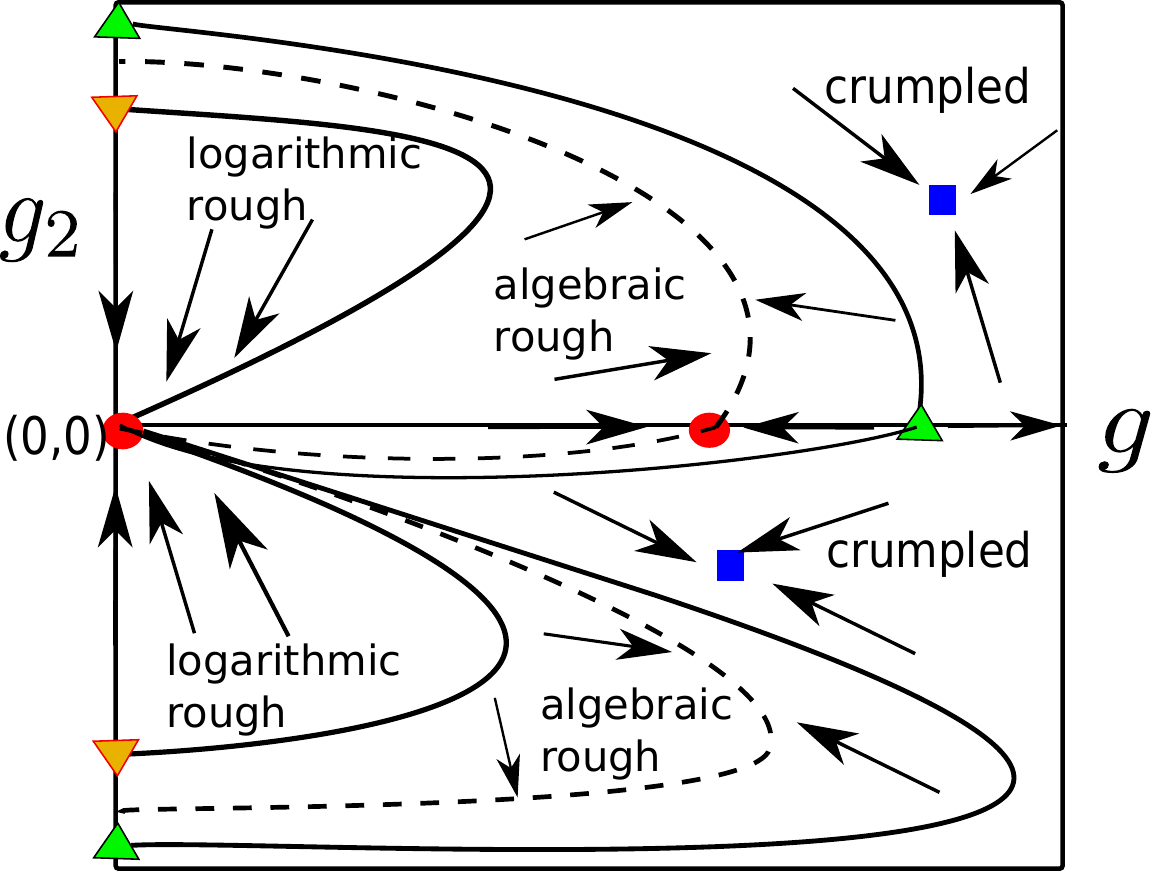}
 \caption{Conjectured “Occam’s razor” global RG flows in the $g$-$g_2$ plane. Different speculated fixed points/lines characterizing the rough and crumpled phases are marked. The broken lines are speculated, fixed lines in the rough phase; see text.}\label{occam}
\end{figure}

We now focus on the region where $0<\chi<1$. When $B=0$, which gives the separatrices (\ref{sep-eq1}), we have $\chi=0$, giving two solutions for $\gamma=-1.383,\,0.161$. These are the red lines in Fig.~\ref{flow-2d}(a). Within the region demarcated by $0<\chi<1$, $\chi$ follows Eq.~(\ref{chi-val-2d}), which is parametrized by $\gamma$. Thus we have nonuniversal scaling exponents, which have their origin in the competition between the nonlocal and local nonlinearities in \eqref{ch-kpz}. 
 The variation of $\chi$ with $\gamma$ is depicted in Fig.~\ref{flow-2d}(c). In a RG language, this nonuniversality in scaling indicates a {\em fixed line}, as opposed to a fixed point, that characterizes the rough phase. Such nonuniversal $\chi(\gamma)$ are reminiscent of the analogous nonuniversal exponents found in the logarithmically rough phase in Ref.~\cite{active_kpz}. 
{ The vanishing of $\chi$ at $\gamma=\gamma_{c2}=-1.383$ and $\gamma=\gamma_{c1}=0.161$  defines the boundary between regions with logarithmically rough and  algebraically rough surfaces, which exactly matches with the predictions from RG analysis~\cite{active_kpz}.} Outside the unstable region, i.e., for $\gamma>\gamma_{c1}$ or $\gamma<\gamma_{c2}$, one has $\chi<0$, which falls outside the validity of our MCT. Since $\chi<0$ is predicted from (\ref{chi-val-2d}) for $|\gamma|\rightarrow \infty$, on the $g_2$ axis we have the logarithmically rough phase found in Ref.~\cite{active_kpz}.  At $\gamma=0$ we recover $\chi=1/3$ for the 2D KPZ equation~\cite{jkb-mct}. 

Having ascertained the existence of rough and crumpled phases in 2D, we now speculate on the global fixed point structure, including the strong coupling fixed points, which are inaccessible to perturbative RG; see Fig.~\ref{occam}. The RG flow lines that flow out of the fixed point (0,0) in the unstable region of the phase space should flow to one of these fixed points. To analyze these fixed points, we are first guided by the expectation that there should be a strong coupling fixed point at $g\neq 0, g_2=0$, which is the expected 2D KPZ strong coupling fixed point and should govern the scaling laws for a 2D KPZ surface (which is rough). This is marked by a red circle (the other red circle at the origin is another fixed point that is unstable along the $g$ direction but stable along the $g_2$ direction).  In the present model, this fixed point is expected to be unstable against perturbation by a finite $\gamma$ or $g_2$. On one side of the $g$ axis, there should be a crumpled phase, which we expect to be characterized by a stable crumpled phase fixed point, marked by a blue square in Fig.~\ref{occam}. On both sides of the $g$ axis, there should be a rough phase enclosing the $g$ axis. Considering the MCT prediction of nonuniversal scaling in the rough phase, we speculate a {\em fixed line} in each of these regions marked by broken line in Fig.~\ref{occam}, corresponding to the rough phase. On the side of the $g$ axis with a crumpling fixed point, we also expect to have another region with a stable rough phase fixed line for higher negative values of $g_2$ (or higher negative values of $\gamma$), giving nonuniversal scaling.  Finally, we generally expect that for larger noises, i.e., with larger values of $D$, for which both $g$ and $g_2$ grow, the surface should be less and less ordered. This line of reasoning suggests that there should be two roughening fixed points on the $g_2$ axis on either side of the $g$ axis (marked by inverted triangles), describing roughening of logarithmically rough  crumpling fixed points on the $g$ axis beyond the putative 2D KPZ rough phase fixed point, as well as on the $g_2$ axis (marked by  green triangles), indicating eventual crumpling of the surface for high enough $g$ or $g_2$. These are basically ``Occam's razor'' style arguments: Fig.~\ref{occam} has the simplest topology that naturally reduces to the known RG flow lines for small $g,\,g_2$, as shown in Fig.~\ref{flow-2d}(a). At the same time, it gives the putative global flow lines, allowing for a transition to presumed rough and crumpled phases; see Refs.~\cite{tb2,sm2,sm-mbe} for similar Occam's razor arguments in different contexts.

{We now consider the model in higher dimensions $d>2$.  In $d=2+\epsilon>2$, there are additional fixed points of the RG flows other than (0,0) (stable fixed point) in the $g$-$g_2$ plane. In the accessible region, i.e., $\tilde A(\gamma)>0$, (0,0) is the only stable fixed point and the long-wavelength scaling behavior is identical to that of the Edwards-Wilkinson (EW) equation, characterized by $z=2,\,\chi=(2-d)/2$~\cite{active_kpz}.  In the perturbatively inaccessible strong coupling phase, i.e., $\tilde A(\gamma)<0$, from an RG analysis Ref.~\cite{active_kpz} argued a roughening transition with an associated $\chi=-\epsilon\,\tilde C(\gamma)/{\tilde A(\gamma)}$; $\tilde C(\gamma)\equiv \gamma^2/2+5\gamma/8$. For any $\epsilon>0$, at the roughening transition $\chi$ can grow arbitrarily large, if $\tilde A(\gamma)$ becomes small enough (remaining negative).   As $\chi$ exceeds unity, our theory (\ref{ch-kpz}) breaks down, as   $\langle ({\boldsymbol\nabla}h)^2\rangle$ diverges with $L$, giving a crumpled surface. [As before, it suggests that   as $\chi$ exceeds unity, higher order nonlinear terms not included in (\ref{ch-kpz})
become relevant (in the RG sense), making our theory given by Eq.~\eqref{ch-kpz} to breakdown.] Setting $\chi=1$, we get a limit on $\gamma$, up to which this roughening transition between a smooth phase and an algebraically rough phase can be observed. Beyond this threshold value of $\gamma$, there is a range of $\gamma$ bounded by another threshold such that in the entire range $\tilde A(\gamma)$ is small enough to make $\chi>1$, suggesting that the roughening transition in this range of $\gamma$ is actually between a smooth phase and a crumpled phase (and not a rough phase)! Overall thus, similar to our discussions for 2D, we expect the strong coupling phase to consist of a rough phase with $0<\chi<1$ and a crumpled phase with $\chi>1$. See Fig.~\ref{3d-rg}.
\begin{figure}[htb]
 \includegraphics[width=0.5\columnwidth]{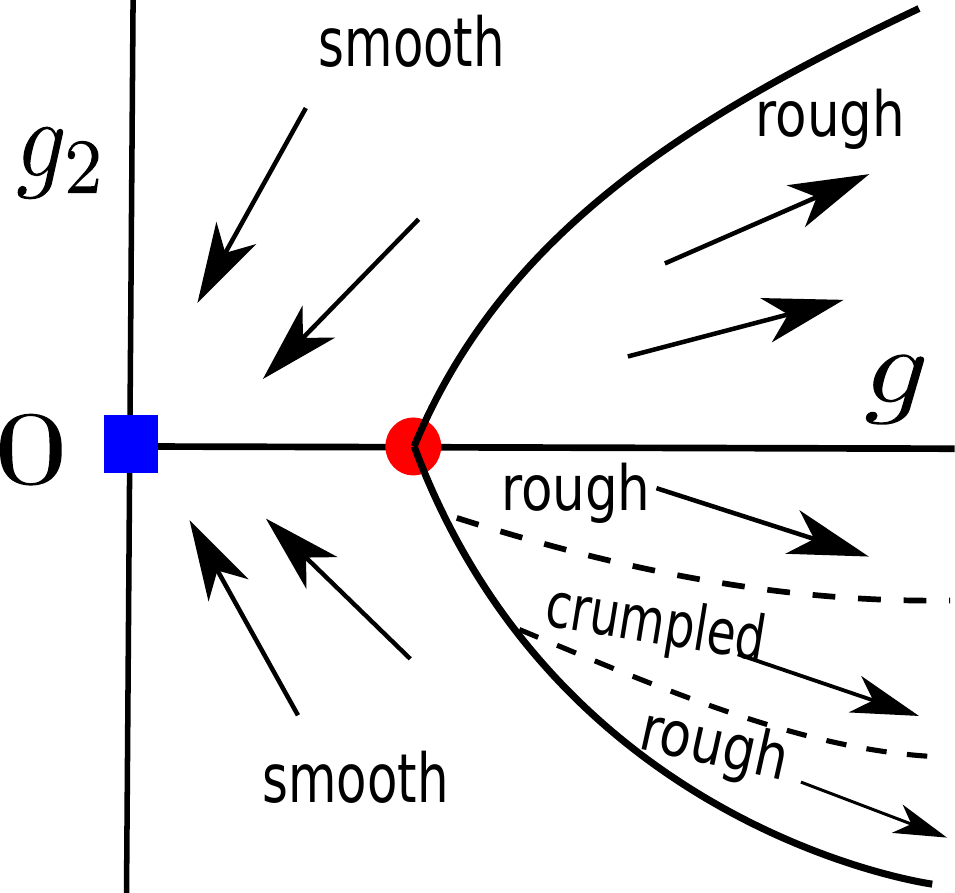}
 \caption{Schematic phase diagram together with RG flow lines in the $g$-$g_2$ plane for $d>2$. The blue square at the origin (0,0) is a stable fixed point that controls the scaling of the smooth phase. The red circle is an unstable fixed point found in the KPZ equation at $d>2$, which demarcates the strong coupling rough phase and the smooth phase. The solid curve lines delineate the strong coupling phases and the smooth phase. Within the strong coupling phase regions, the small region bounded between the broken lines, drawn schematically, correspond to the crumpled phase, the rest being the usual rough phase. Arrows indicate the RG flow directions.}\label{3d-rg}
\end{figure}


We now briefly discuss the MCT predictions for $d>2$ for the strong coupling phases in Fig.~\ref{3d-rg}. Our RG calculations already predicted a smooth phase and a roughening transition presumably to a strong coupling phase. The latter, however, cannot be accessed in the RG calculations. Assuming $\chi>0$, we apply MCT to extract the scaling behavior in the strong coupling phase. Focusing on $d=3$, we get $A=36\gamma^2+48\gamma+15$, $B=54\gamma^2+78\gamma-2.5$, and $C=16\gamma^2+24\gamma$, giving $\chi\equiv \chi(\gamma)$. Thus, both $\chi$ and $z=2-\chi$ are {\em nonuniversal}, as they are in 2D. Secondly, by varying $\gamma$, one can make $\chi$ bigger, eventually exceeding unity. As in 2D, $\chi>1$ should imply crumpling of the surface, for which our theory breaks down. We therefore find that the strong coupling phase in 3D is similar to 2D. Similar analysis indicates analogous behavior in the strong coupling phase for any $d>2$.}


 We have thus studied the strong coupling phases of a generalized KPZ equation in which local and nonlocal nonlinear effects compete (in the scaling sense). Our results from the RG and bare perturbation theory in 2D suggest that the strong coupling phase region in the parameter space  actually is made of two distinct phases, one an algebraically rough phase, with $0<\chi<1$ corresponding to orientational LRO but positional SRO, and another with $\chi>1$, implying a crumpled phase having both positional and orientational SRO. These are corroborated by our one-loop MCT calculations. The crumpled phase, however, cannot be systematically explored by our theory. Our MCT predicts that the algebraic rough phase is characterized by nonuniversal scaling exponents parametrized by $\gamma$, similar to the nonuniversal exponents in the logarithmically rough phase predicted by the RG calculations. Similar behavior including a crumpled phase and a rough phase with nonuniversal scaling exponents are predicted by MCT in 3D. The results in Ref.~\cite{active_kpz} showed that chirality has the effect of suppressing the nonlinear instabilities of the RG flow. We thus expect that chiral effects together with the nonlocality should be able to suppress the crumpling of the membrane, stabilizing orientational order.  We note that the nonuniversal scaling exponents found in the MCT calculations are a crucial outcome of the lack of vertex renormalization in one-loop MCT. We are unable to speculate whether this result is protected by any ``hidden'' symmetry not known at present, or a fortuitous result, or will not hold at higher order perturbation theory. Indeed, if there are unequal infinite renormalizations of $\lambda,\,\lambda_1$ at some higher loop order, their ratio $\gamma$ will acquire nontrivial fixed point value(s), instead of being marginal. In this case, a rough phase would be characterized by isolated fixed point(s), which should replace the fixed lines in the Occam's razor global RG flow diagram (\ref{occam}). Thus, it would be interesting to verify our results numerically, e.g., by using pseudospectral methods~\cite{bmhd2,ab-mhd} or by nonperturbative methods~\cite{nonpert3,nonpert2,nonpert1}. However, we expect our predictions on the existence of a crumpled phase to hold true. Our theory should apply to nonequilibrium surfaces with other fast relaxing active degrees of freedom living on it. We look for future experimental endeavors along this direction.
 
 {\em Acknowledgement.} A.B. thanks 
the Alexander von Humboldt Stiftung, Germany for partial financial support through the Research Group Linkage Programme (2024). The authors thank T. Halpin-Healy for various helpful suggestions.

\bibliography{MCT.bib}

\clearpage
\onecolumngrid
\begin{center}
\textbf{\large Supplemental Material for ``Rough or crumpled: Strong coupling phases of a generalized Kardar-Parisi-Zhang surface''}
\end{center}
\vspace{0.5cm}

\section{Galilean invariance}
We rewrite the generalized KPZ equation  Eq.~(2) in the main text, which was introduced as a conceptual model incorporating competing local and nonlocal nonlinear interactions,
\begin{align}
 \frac{\partial h}{\partial t} &= \nu \nabla^2 h+\frac{\lambda}{2} (\boldsymbol\nabla h)^2 +  \lambda_1 \int d^dr' Q_{ij}({\bf r-r'})(\nabla_ih ({\bf r'},t)\nabla_jh({\bf r'},t)) + \eta, 
 \label{ch-kpz_sup}
\end{align}
where $Q_{ij}({\bf x})\equiv \nabla_i\nabla_j/\nabla^2$ is the longitudinal projection operator and takes the form $k_ik_j/k^2$ in the Fourier space; $\bf k$ is a Fourier wavevector.
Under the transformation $x_i'\rightarrow x_i-(\lambda+2\lambda_1)c_it$ and $t'\rightarrow t$, where $i$ and $j$ can take values $1$ or $2$ we obtain,
\begin{align}
&\frac{\partial}{\partial t'}\rightarrow \frac{\partial}{\partial t}+(\lambda+2\lambda_1)\textbf{c}\cdot\boldsymbol\nabla;\quad\boldsymbol\nabla'\rightarrow \boldsymbol\nabla.
\end{align} 
Using these we show below that if the height function $h({\bf x},t)$ transforms as $h'({\bf x}',t')\rightarrow h({\bf x},t)+\textbf{c}\cdot\textbf{x}$, then Eq.~(\ref{ch-kpz_sup}) remains  invariant. Considering RHS of Eq.~(\ref{ch-kpz_sup}), the first term transforms as
\begin{align}
\nu \nabla'^2 h'\rightarrow \nu \nabla^2 h.
\end{align} 
The second term transforms as
\begin{align}
\frac{\lambda}{2} (\boldsymbol\nabla' h')^2\rightarrow \frac{\lambda}{2} (\boldsymbol\nabla h)^2+\lambda \boldsymbol \nabla h \cdot \textbf{c}+\frac{\lambda}{2}c^2.
\end{align} 
The third term transforms as
\begin{align}
\lambda_1 \frac{\nabla'_i \nabla'_j}{\nabla'^2}(\nabla'_ih' \nabla'_jh')\rightarrow \lambda_1 \frac{\nabla_i \nabla_j}{\nabla^2}(\nabla_ih \nabla_jh)+2\lambda_1 \textbf{c}\cdot\boldsymbol\nabla h.
\end{align}
Noise term remains invariant. Similarly LHS of Eq.~(\ref{ch-kpz_sup}) transforms as
\begin{align}
 \frac{\partial h'}{\partial t'}\rightarrow \frac{\partial h}{\partial t}+(\lambda+2\lambda_1)\textbf{c}\cdot\boldsymbol\nabla h+(\lambda+2\lambda_1)c^2.
\end{align} 
From the above equations we see that Eq.~(\ref{ch-kpz_sup}) is invariant.


\section{Renormalization group calculations}

We revisit and reanalyze the renormalization group calculations on the generalized KPZ equation (2) of the main text. We follow~\cite{active_kpz}. We first give the path integral over $h({\bf r},t)$ and its dynamic conjugate field $\hat h({\bf r},t)$~\cite{bausch,tauber} that is equivalent to and constructed from Eq.~(2) of the main text together with the noise variance. 
The generating functional corresponding to Eq.~(2) of the main text is given by~\cite{bausch,tauber}
\begin{equation}
 \mathcal{Z}=\int \mathcal{D}\hat{h} \mathcal{D}h e^{-\mathcal{S}[\hat{h},h]},
\end{equation}
where $\hat{h}$ is the dynamic conjugate field and $\mathcal{S}$ is the action functional:
\begin{align} 
{\mathcal S}= -\int_{{\bf x},t}\hat{h}D\hat{h} + \int_{{\bf x},t}\hat{h}\bigg\{\partial_th-\nu \nabla^2h-\frac{\lambda}{2} (\boldsymbol{\nabla} h)^2-\lambda_1 \frac{\nabla_i \nabla_j}{\nabla^2}(\nabla_ih \nabla_jh)\bigg\}. \label{action_sup}
\end{align}
Using the one loop diagrams from the supplemental material of Ref.~\cite{active_kpz} we obtain renormalized $D$ ($D^<$) and renormalized $\nu$ ($\nu^<$), we get
\begin{align}
D^{<}=D \Biggr[ 1+\frac{\lambda^2D}{\nu^3}k_d\biggl(\frac{1}{4}+\gamma\frac{1}{d}+\gamma^2\frac{3}{d(d+2)}\biggl)\int_{\frac{\Lambda}{b}}^{\Lambda}\frac{d^dq}{q^2} \Biggr],
\end{align}
\begin{align}
\nu^{<}=\nu \Biggr[ 1+\frac{\lambda^2D}{\nu^3}k_d\biggl(\frac{2-d}{4d}+\frac{\gamma}{d(d+2)}\Bigl(3-\frac{d+2}{2}+\frac{d(d+2)}{2}\Bigl)+\frac{\gamma^2}{d}\biggl)\int_{\frac{\Lambda}{b}}^{\Lambda} \frac{d^dq}{q^2} \Biggr].
\end{align} 
Here $k_d=\frac{S_d}{(2\pi)^d}$, where $S_d$ is the surface area of a $d$-dimensional unit sphere and $\gamma=\frac{\lambda_1}{\lambda}$ as defined in the main text. Using above equations and Setting $d=2$ in the above integrals, with $b=e^{\delta l}\approx 1+\delta l$ ($l$ is the RG time) and defining dimensionless coupling constant by $g\equiv\frac{\lambda^2D}{\nu^3}k_d$ as in the main text, we obtain the RG flow equations for $D$ and $\nu$
\begin{align}
\frac{dD}{dl}=D \Biggr[z-d-2\chi+g\biggl(\frac{1}{4}+\gamma\frac{1}{d}+\gamma^2\frac{3}{d(d+2)}\biggl)\Biggr],
\end{align}
\begin{align}
\frac{d\nu}{dl}=\nu \Biggr[z-2+g\biggl(\frac{2-d}{4d}+\frac{\gamma}{d(d+2)}\Bigl(3-\frac{d+2}{2}+\frac{d(d+2)}{2}\Bigl)+\frac{\gamma^2}{d}\biggl) \Biggr].
\end{align}
Furthermore, defining another dimensionless effective coupling constant, $g_1\equiv \frac{\lambda_1^2 D}{\nu^3}k_d,\,g_2\equiv \text{sign}(\gamma)\sqrt{g_1}$ and using above equations we can write the flow equations of $g$ and $g_2$:
\begin{align*}
\frac{dg}{dl}=g\Biggr[2-d+g\biggl\{ \frac{1}{4}+\frac{3}{d(d+2)}\gamma^2+\frac{\gamma}{d}-\frac{3}{4}\frac{2-d}{d}-\frac{3}{d}\gamma^2-\frac{3\gamma}{d(d+2)}\biggl(3-\frac{d+2}{2}+\frac{d(d+2)}{2}\biggl) \biggl\}\Biggr],
\end{align*} 
\begin{align*}
\frac{dg_2}{dl}=\frac{g_2}{2}\Biggr[2-d+g\biggl\{ \frac{1}{4}+\frac{3}{d(d+2)}\gamma^2+\frac{\gamma}{d}-\frac{3}{4}\frac{2-d}{d}-\frac{3}{d}\gamma^2-\frac{3\gamma}{d(d+2)}\biggl(3-\frac{d+2}{2}+\frac{d(d+2)}{2}\biggl) \biggl\}\Biggr].
\end{align*} 
To the lowest order in $d-d_c$, where $d_c=2$ is the critical dimension,
\begin{eqnarray}
 \frac{dg}{dl}&=&g[2-d-g\tilde A(\gamma)],\label{flow-g_sup}\\
  \frac{dg_2}{dl}&=&\frac{g_2}{2}[2-d-g\tilde A(\gamma)],\label{flow-g*_sup}
\end{eqnarray}
where $\tilde A(\gamma)=\frac{9}{8}\gamma^2+\frac{11}{8}\gamma-\frac{1}{4}$ as given in the main text.

\section{MCT Calculations}

We start with the general definitions of the correlation function $C({ {\bf k}},\omega)$ and response function $G({ {\bf k}},\omega)$:
\begin{align}
&\delta^{(d)}({\bf k+k^{'}})\delta(\omega+\omega^{'})G({ {\bf k}},\omega)=\biggl\langle\frac{\delta h({\bf k},\omega)}{\delta \eta({\bf k^{'}},\omega^{'})}\biggr\rangle\\
&\delta^{(d)}({\bf k+k^{'}})\delta(\omega+\omega^{'})C({ {\bf k}},\omega)=\langle h({\bf k},\omega)h({\bf k^{'}},\omega^{'})\rangle.
\end{align}
The response and correlation functions are assumed to have the following scaling form in the long time large length scale limit:
\begin{align}
&G({\bf k},\omega)=k^{-z}g\biggl(\frac{\omega}{k^z}\biggr)\\
&C({\bf k},\omega)=k^{-d-2\chi-z}f\biggl(\frac{\omega}{k^z}\biggr),
\end{align}
where $z$ is the dynamic exponent and $\chi$ is the roughness exponent; see also main text for their definitions.
The Dyson's equation for the self energy $\Sigma$ in the scaling limit is 
\begin{align}
G^{-1}({\bf k},\omega)=-i\omega+\Sigma({\bf k},\omega).
\end{align}
The zero frequency self energy or the relaxation rate has the form
\begin{align}
\Sigma({\bf k},0)=\Gamma k^z. \label{selfeng_1_sup}
\end{align}
Here $\Gamma>0$ is the effective damping coefficient. The correlation function in Lorentzian approximation can be written as 
\begin{align}
C({\bf k},\omega)=\frac{2D\Gamma k^{-2\chi-d+z}}{\omega^2+\Gamma^2k^{2z}}.
\end{align}
Furthermore, the zero-frequency correlation function is
\begin{align}
C({\bf k},0)=\frac{2D}{\Gamma} k^{-2\chi-d-z}.\label{correl_1_sup}
\end{align}
\begin{figure}[htb]
\includegraphics[width=0.44\textwidth]{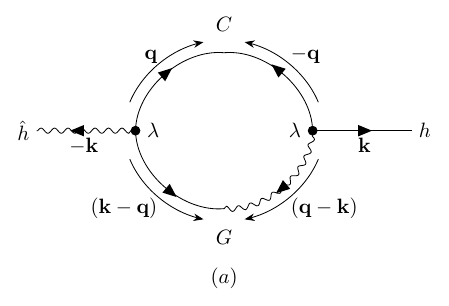}
\includegraphics[width=0.44\textwidth]{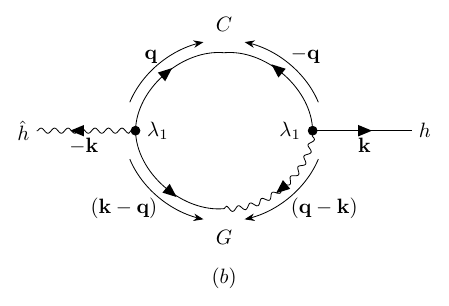}\\
\includegraphics[width=0.44\textwidth]{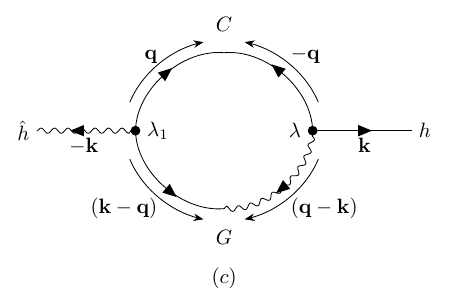}
\includegraphics[width=0.44\textwidth]{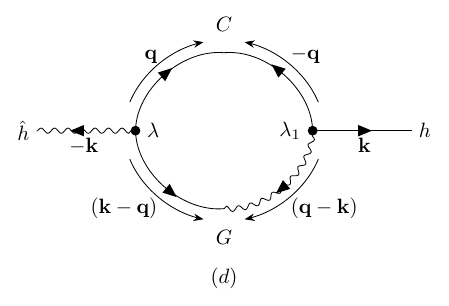}
\caption{ One-loop Feynman diagrams that contribute to the zero frequency self energy.}
\label{self_eng_diag}
\end{figure}
Figure~\ref{self_eng_diag}(a) shows the one-loop diagrammatic correction to the self energy at ${\cal O}(\lambda^2)$ that comes with a symmetry factor of $8$ and contributes 
\begin{align}
&\Bigl(\frac{\lambda}{2}\Bigl)^2\frac{1}{2!}\, 8\int_{{\bf q},\Omega}{\bf q\cdot(k-q)(q\cdot k)}C({\bf q}, \Omega)G({\bf k-q},-\Omega)
=\lambda^2\int_{{\bf q},\Omega}\frac{{\bf (k-q)\cdot q\times(k\cdot q)}\times 2D\Gamma q^{-2\chi-d+z}}{(\Omega^2+\Gamma^2q^{2z})(i\Omega+\Gamma({\bf k-q})^z)}.
\end{align}
After performing the $\Omega$-integral and making ${\bf q}\rightarrow{ ({\bf q}+\frac{\bf k}{2})}$, above integral becomes
\begin{align}
I&=\frac{\lambda^2D}{\Gamma}\int_{\bf q}\Bigl({\bf q}+\frac{{\bf k}}{2}\Bigl)^{-2\chi-d+z} { \biggl[\Bigl(\frac{\bf k}{2}-{\bf q}\Bigl)\cdot\Bigl|{\bf q}+\frac{\bf k}{2}\Bigl|\biggl]}\frac{{ \biggl[{\bf k}\cdot \Bigl|{\bf q}+\frac{\bf k}{2}\Bigl|\biggl]}}{{ \Bigl|{\bf q}+\frac{\bf k}{2}\Bigl|}^z+{ \Bigl|{\bf q}-\frac{\bf k}{2}\Bigl|}^z}\nonumber \\&=\frac{\lambda^2D}{2\Gamma}\frac{2\chi+d}{2}\int_{\bf q}\frac{q^2({\bf k\cdot q})q^{-2\chi-d}}{q^z}\frac{({\bf k\cdot q})}{q^2}-\frac{\lambda^2D}{2\Gamma}k^2\int_{\bf q} \frac{q^2}{2q^z}q^{-2\chi-d}.
\end{align}
After simplifications, the first contribution in the second line of the above integral becomes
\begin{align*}
\frac{\lambda^2D}{2\Gamma}\,\frac{2\chi+d}{2}\,\frac{k_d}{d}\times k^2\int_{\bf q}q^{-\chi-1}.
\end{align*}
Similarly, the second contribution gives
\begin{align*}
\frac{\lambda^2D}{2\Gamma}\,\frac{k_d}{2}\, k^2\int_{\bf q}q^{-\chi-1}.
\end{align*}
Adding the above contributions, we get the ${\cal O}(\lambda^2)$ correction to the zero frequency self energy
\begin{align}
I=\frac{\lambda^2D}{4\Gamma}\frac{2}{d}k_dk^{2-\chi}.\label{s_e_4}
\end{align}


Figure~\ref{self_eng_diag}(b) shows the one-loop contribution to the self energy at ${\cal O}(\lambda_1^2)$ with a symmetry factor of $8$ and contributes
\begin{align}
I&=\frac{\lambda_1^2}{2!}\, 8\int_{{\bf q},\Omega}\frac{k_ik_j}{k^2}q_i(k-q)_j\frac{(q-k)_m(q-k)_n}{(q-k)^2}q_mk_n C({\bf q},\Omega)\times G({\bf k-q},-\Omega)\nonumber\\&=\frac{4\lambda_1^2k_ik_jk_n}{k^2}\int_{{\bf q},\Omega}q_iq_m(k-q)_j(q-k)_m(q-k)_n\frac{1}{{\bf (q-k)^2}}\times\frac{2D\Gamma q^{-2\chi-d+z}}{(\Omega^2+\Gamma^2q^{2z})}\times\frac{1}{i\Omega+\Gamma{\bf |k-q|}^z}.
\end{align}
After performing the $\Omega$-integral,  we get
\begin{align}
&I=\frac{4\lambda_1^2D}{\Gamma}\frac{k_ik_jk_n}{k^2}\int_{\bf q}\frac{q_iq_m(k-q)_j(q-k)_m(q-k)_nq^{-2\chi-d}}{{\bf (q-k)}^2(q^z+{\bf |k-q|}^z)}.
\end{align}
Then shifting ${\bf q}\rightarrow{\bf q}+\frac{\bf k}{2}$, above integral becomes
\begin{align}
&\frac{4\lambda_1^2D}{\Gamma}\frac{k_ik_jk_n}{k^2}\int_{\bf q}\frac{(q+\frac{k}{2})_i(\frac{k}{2}-q)_j(q-\frac{k}{2})_m(q-\frac{k}{2})_n(q+\frac{k}{2})_m}{{({\bf q}-\frac{{\bf k}}{2})}^2({|{\bf q}+\frac{\bf k}{2}|}^z+{ |{\bf q}-\frac{\bf k}{2}|}^z)}{|{\bf q}+\frac{\bf k}{2}|}^{-2\chi-d}.
\end{align}
Next, we extract the ${\cal O}(k)$ and ${\cal O}(k^0)$ contributions from the numerator. Then above integral $I$ can be split into three parts $I_1$, $I_2$ and $I_3$, with
\begin{align}
I_1&=-\frac{4\lambda_1^2D}{2\Gamma}\frac{k_ik_jk_mk_n}{k^2}\int_{\bf q} q_iq_jq_mq_nq^{2-2\chi-d-z-4}=-\frac{6\lambda_1^2D}{\Gamma d(d+2)}\frac{1}{\chi}k_dk^{-\chi+2},
\end{align}
\begin{align}
I_2&=\frac{4\lambda_1^2D}{\Gamma}\frac{k_ik_jk_mk_n}{k^2}\int_{\bf q} \frac{q_iq_jq_mq_nq^{-2\chi-d}}{2q^2q^z}\times \frac{2\chi+d}{2}=\frac{3\lambda_1^2D}{\Gamma d(d+2)}\frac{2\chi+d}{\chi}k_dk^{-\chi+2},
\end{align}
where we have used the following identity~\cite{yakhot}, 
\begin{align} 
&k_ik_jk_mk_n\int d^dqf(q^2)q_iq_jq_mq_n=k_ik_jk_mk_n\frac{[\delta_{ij}\delta_{mn}+\delta_{im}\delta_{jn}+\delta_{in}\delta_{jm}]}{d(d+2)}\int d^dq f(q^2)q^4\label{identity_2},
\end{align}
and
\begin{align}
I_3&=\frac{4\lambda_1^2D}{\Gamma}k_ik_j\int_{\bf q} \frac{q_iq_jq^2q^{-2\chi-d}}{4q^2q^z}=\frac{\lambda_1^2D}{\Gamma d}\frac{1}{\chi}k_dk^{-\chi+2}.
\end{align}
Then,
$I_3$ is evaluated by using the well-known relation~\cite{yakhot}
\begin{align} 
k_ik_j\int d^dqf(q^2)q_iq_j=k_ik_j\times\frac{[\delta_{ij}]}{d}\int d^dq f(q^2)q^2.\label{identity_1}
\end{align}
Adding $I_1$, $I_2$ and $I_3$ we get contribution to zero frequency self energy at ${\cal O}(\lambda_1^2)$:
\begin{align}
-\frac{\lambda_1^2 D}{\Gamma}\frac{k_d}{d}\frac{1}{\chi}k^{2-\chi}\biggl(\frac{4-4d-6\chi}{d+2}\biggl).\label{s_e_1}
\end{align}

Figure~\ref{self_eng_diag}(c) gives one of the two one-loop corrections to the  self energy at  ${\cal O}(\lambda\lambda_1)$, 
with a contribution
\begin{align}
&I=\frac{4\lambda \lambda_1}{2!}\frac{k_ik_j}{k^2}\int_{{\bf q},\Omega}q_i(k-q)_j({\bf q\cdot k})\frac{1}{i\Omega+\Gamma{ \bf |k-q|}^z} \frac{2D\Gamma q^{-2\chi-d+z}}{\Omega^2+\Gamma^2q^{2z}}. 
\end{align} 
After performing the $\Omega$-integral and shifting ${\bf q}\rightarrow{\bf q}+\frac{\bf k}{2}$, we get
\begin{align}
&\frac{2\lambda \lambda_1 D}{\Gamma}\frac{k_ik_jk_m}{k^2}\int_{{\bf q}}(q+\frac{k}{2})_i(\frac{k}{2}-q)_j(q+\frac{k}{2})_m\frac{|{\bf q}+\frac{\bf k}{2}|^{-2\chi-d}}{{|{\bf q}+\frac{\bf k}{2}|}^z+| \frac{\bf k}{2}-{\bf q}|^z}.
\end{align}

\begin{align}
&=-\frac{2\lambda \lambda_1 D}{\Gamma}\times\frac{k_ik_jk_m}{k^2}\int_{{\bf q}}q_iq_j\bigl(q_m+\frac{k_m}{2}\bigl)\frac{q^{-2\chi-d}}{2q^z} \biggl(1+\frac{{\bf k\cdot q}}{q^2}\biggl)^{-(\frac{2\chi+d}{2})}\\&=\frac{2\lambda \lambda_1 D}{\Gamma}\frac{k_ik_jk_mk_n}{k^2}\frac{2\chi+d}{2}\int_{{\bf q}}\frac{q_iq_jq_mq_n}{2q^2q^z}q^{-2\chi-d}-\frac{2\lambda \lambda_1 D}{\Gamma}\frac{k_ik_j}{2}\int_{{\bf q}}\frac{q_iq_j}{2q^z}q^{-2\chi-d}\\&=\frac{\lambda \lambda_1 D}{\Gamma}3k^2k_d\frac{2\chi+d}{2}\frac{1}{d(d+2)}\int_{{\bf q}}q^{4-2\chi-d-z-2+d-1}-\frac{\lambda \lambda_1 D}{2\Gamma}\frac{k^2k_d}{d}\int_{{\bf q}}q^{2-2\chi-d-z+d-1}.
\end{align}
In evaluation of these integrals we have used identities~(\ref{identity_2}) and~(\ref{identity_1}).
After some simplifications, the above integral gives
\begin{align}
-\frac{\lambda \lambda_1 D}{\Gamma}\frac{k_d}{2d}\frac{k^{2-\chi}}{\chi}\biggl(\frac{2-6\chi-2d}{d+2} \biggl).\label{s_e_2}
\end{align}
Finally,
Fig.~\ref{self_eng_diag}(d) gives the second one-loop correction to the self energy that is also ${\cal O}(\lambda\lambda_1)$, but is distinct from the one in Fig.~\ref{self_eng_diag}(c).
\begin{align}
I=\frac{4\lambda \lambda_1}{2!}\int_{{\bf q},\Omega}{\bf (q\cdot (k-q))}\frac{(q-k)_i(q-k)_j}{{\bf (q-k)}^2}q_ik_j\times \frac{1}{i\Omega+\Gamma{ \bf |k-q|}^z}\times\frac{2D\Gamma q^{-2\chi-d+z}}{\Omega^2+\Gamma^2q^{2z}}. 
\end{align}
After performing the $\Omega$-integral, above integral becomes
\begin{align}
\frac{2\lambda \lambda_1 D}{\Gamma}k_j\int_{{\bf q}}\frac{{\bf (q\cdot (k-q))}(q-k)_i(q-k)_jq_iq^{-2\chi-d}}{{\bf (q-k)}^2\times(q^z+{\bf |k-q|}^z)}.
\end{align}
After shifting ${\bf q}\rightarrow{\bf q}+\frac{\bf k}{2}$, above integral reduces to
\begin{align}
\frac{2\lambda \lambda_1 D}{\Gamma}k_j\int_{{\bf q}}(q+\frac{k}{2})_m(\frac{k}{2}-q)_m(q-\frac{k}{2})_i(q-\frac{k}{2})_j(q+\frac{k}{2})_i\times\frac{{\bf (q+\frac{k}{2})}^{-2\chi-d}}{{({{\bf q}-\frac{\bf k}{2}})^2\biggl(|{\bf q}+\frac{\bf k}{2}}|^z+|{\bf q}-\frac{\bf k}{2}|^z\biggl)}.
\end{align}
Next we calculate the ${\cal O}(k)$ and ${\cal O}(k^0)$ contributions from numerator, which are relevant. Then the above integral $I$ can be split into three parts say $I_1$, $I_2$ and $I_3$. The first part $I_1$, after expanding its denominator binomially, is
\begin{align}
I_1&=-\frac{2\lambda\lambda_1 D}{\Gamma}k_j\int_{{\bf q}}\frac{q^4q_jq^{-2\chi-d}}{2q^z}\,\frac{k_iq_i}{q^4}=-\frac{\lambda\lambda_1 D}{\Gamma}\frac{k_d}{d}\frac{k^{2-\chi}}{\chi}.
\end{align}
Next,
\begin{align}
I_2&=\frac{2\lambda\lambda_1 D}{\Gamma}k_j\int_{{\bf q}}\frac{q^4q_jq^{-2\chi-d}}{q^2\times 2q^z}\times\frac{2\chi+d}{2}\times\frac{k_iq_i}{q^2}=\frac{\lambda\lambda_1 D}{2\Gamma}
(2\chi+d)\frac{k_d}{d}\frac{k^{2-\chi}}{\chi}.
\end{align}
In evaluating $I_1$ and $I_2$ we used relation~(\ref{identity_1}).
Finally,
\begin{align}
I_3&=\frac{2\lambda\lambda_1 D}{2\Gamma}k^2\int_{{\bf q}}\frac{q^4q^{-2\chi-d}}{2q^2q^z}=\frac{\lambda\lambda_1 D}{2\Gamma}
\frac{k_dk^{2-\chi}}{\chi}.
\end{align}
Adding above three contributions we get
\begin{align}
{ I_1+I_2+I_3}=-\frac{\lambda\lambda_1 D}{\Gamma}\frac{k_d}{d}\frac{k^{2-\chi}}{\chi}\biggl[\frac{2-2\chi-2d}{2}\biggl].\label{s_e_3}
\end{align}
Adding Eqs.~(\ref{s_e_2}) and~(\ref{s_e_3})  we get the total ${\cal O}(\lambda\lambda_1)$ contribution to  self energy at zero-frequency
\begin{align}
\frac{\lambda^2D}{\Gamma}\gamma\frac{k_d}{2d}\frac{k^{2-\chi}}{\chi}\biggl[\frac{4d+2d^2+2\chi d+10\chi -6}{d+2}\biggl].\label{s_e_5}
\end{align}
Here, $\gamma=\frac{\lambda_1}{\lambda}$, as originally defined in the main text. Now adding Eqs.~(\ref{s_e_4}),~(\ref{s_e_1}) and~(\ref{s_e_5}) we get total one-loop contribution to the zero frequency self energy. Now using the definition given in Eq.~(\ref{selfeng_1_sup}), we obtain the relation 
\begin{align}
\frac{\Gamma^2}{\lambda^2D}=\frac{k_d}{\chi\, d}\biggl[\frac{\chi}{2}+\gamma\biggl(\frac{4d+2d^2+2\chi d+10\chi-6}{2(d+2)}\biggl)+\gamma^2\biggl(\frac{6\chi+4d-4}{d+2}\biggl)\biggl].\label{relation_1_sup}
\end{align}

\begin{figure}[htb]
\includegraphics[width=0.44\textwidth]{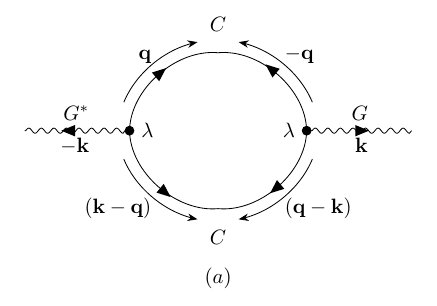}
\includegraphics[width=0.44\textwidth]{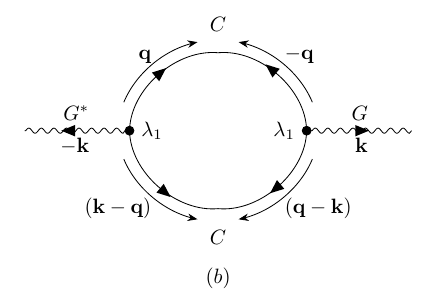}
\includegraphics[width=0.44\textwidth]{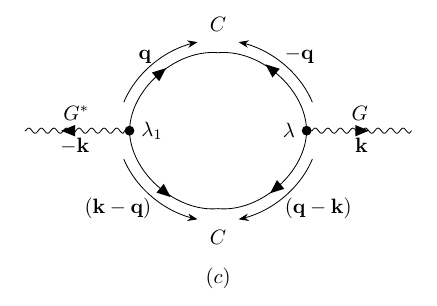}
\caption{ One-loop Feynman diagrams that contribute to the zero frequency correlation function.}
\label{corr_func_diag}
\end{figure}
Next we calculate the one-loop contributions to the correlation function; see Fig.~\ref{corr_func_diag}.

Figure~\ref{corr_func_diag}(a) gives the one-loop correction at ${\cal O}(\lambda^2)$ to the correlation function with a symmetry factor of $2$. Evaluating it at zero frequency, we get
 \begin{align}
 &2\times\Bigl(\frac{\lambda}{2}\Bigl)^2\frac{1}{2!}\times 2\times\frac{1}{\Gamma^2k^{2z}}\int_{{\bf q},\Omega}\bigl({\bf  q}\cdot({\bf k-q})\bigl)^2 C({\bf q},\Omega)\times C({\bf k-q},-\Omega)\nonumber\\&=\frac{\lambda^2}{2\Gamma^2k^{2z}}\int_{{\bf q},\Omega}\bigl({\bf  q}\cdot({\bf k-q})\bigl)^24D^2\Gamma^2q^{-2\chi-d+z}\times\frac{|{\bf k-q}|^{-2\chi-d+z}}{(\Omega^2+\Gamma^2q^{2z})(\Omega^2+\Gamma^2|{\bf k-q}|^{2z})}\\&=\frac{2D^2\lambda^2}{k^{2z}}\int_{{\bf q}, \Omega}\frac{q^4\times q^{-4\chi-2d+2z}}{(\Omega^2+\Gamma^2q^{2z})^2}
 \end{align}
After performing the $\Omega$-integral, we find
\begin{align}
&\frac{2D^2\lambda^2}{k^{2z}}\times\frac{1}{4\Gamma^3}k_d\int q^{4-4\chi-2d-z+d-1}dq =\frac{\lambda^2D^2}{2\Gamma^3}k_d\frac{1}{(d+3\chi-2)}k^{-d-\chi-2}.\label{c_f_1}
\end{align}
Next, 
Fig.~\ref{corr_func_diag}(b) is the one-loop correction at  ${\cal O}(\lambda_1^2)$ with a symmetry factor of $2$. Evaluating it at zero frequency, we get
\begin{align}
-2\times\frac{\lambda_1^2}{2!}\times2\frac{1}{\Gamma^2k^{2z}}\frac{k_ik_jk_mk_n}{k^4}\int_{{\bf q},\Omega}q_i(k-q)_jq_m(q-k)_n\times\frac{4D^2\Gamma^2q^{-4\chi-2d+2z}}{\bigl(\Omega^2+\Gamma^2q^{2z}\bigl)^2}.
\end{align} 
Using, $\int_\Omega\frac{1}{\bigl(\Omega^2+\Gamma^2q^{2z}\bigl)^2}=\frac{1}{4\Gamma^3q^{3z}}$, above integral reduces to
\begin{align}
\frac{8\lambda_1^2D^2\Gamma^2}{4\Gamma^5k^{2z}}\times\frac{k_ik_jk_mk_n}{k^4}\int_{{\bf q}}q_iq_jq_mq_nq^{-4\chi-2d-z}.
\end{align}
By using $\chi+z=2$ and identity~(\ref{identity_2}), we get the one-loop contribution to the correlation function at ${\cal O} (\lambda_1^2)$:
\begin{align}
\frac{6\lambda^2D^2}{\Gamma^3}\gamma^2\frac{k_d}{d(d+2)}\frac{1}{(d+3\chi-2)}k^{-d-\chi-2}.\label{c_f_2}
\end{align}
Finally,
Fig.~\ref{corr_func_diag}(c) gives the one-loop contribution to zero frequency correlation function at ${\cal O} (\lambda\lambda_1)$ with a symmetry factor of $2$. Evaluating, we find
\begin{align}
&-2\times\frac{\lambda\lambda_1}{2!}\times2\frac{1}{\Gamma^2k^{2z}}\frac{k_ik_j}{k^2}\int_{{\bf q},\Omega}q_i(k-q)_j\bigl[{\bf q\cdot (q-k)}\bigl]\times\frac{4D^2\Gamma^2q^{-4\chi-2d+2z}}{\bigl(\Omega^2+\Gamma^2q^{2z}\bigl)^2}\\&=\frac{8\lambda\lambda_1D^2\Gamma^2}{4\Gamma^5k^{2z}}\times\frac{k_ik_j}{k^2}\int_{{\bf q}}q_iq_jq^{2-4\chi-2d-z}\\&=\frac{2\lambda_1\lambda D^2}{\Gamma^3k^{2z}}\frac{k_d}{d}\int_q q^{4-4\chi-2d-z+d-1}.
\end{align}
In evaluation of the above integral we used relation $\chi+z=2$ and identity~(\ref{identity_1}). After some simplifications, we finally obtain 
\begin{align}
\frac{2\lambda^2 D^2}{\Gamma^3}\gamma\frac{k_d}{d}\frac{k^{-d-\chi-2}}{(d+3\chi-2)}.\label{c_f_3}
\end{align}
Now, adding Eqs.~(\ref{c_f_1}),~(\ref{c_f_2}) and~(\ref{c_f_3}), we obtain the total contribution to zero frequency correlation function from all combination of vertices at one loop order, which reads
\begin{align}
\frac{\lambda^2D^2}{\Gamma^3}\frac{k_d}{(d+3\chi-2)}k^{-d-\chi-2}\biggl[\frac{1}{2}+\frac{2\gamma}{d}+\frac{6\gamma^2}{d(d+2)}\biggl].\label{c_f_total}
\end{align}
Then equating Eq.~(\ref{c_f_total}) with the definition of the zero frequency correlation function given by Eq.~(\ref{correl_1_sup}) we obtain
\begin{align}
\frac{\Gamma^2}{\lambda^2D}=\frac{k_d}{4d(d+2)(d+3\chi-2)}\biggl[d(d+2)+4\gamma(d+2)+12\gamma^2 \biggl].\label{relation_2_sup}
\end{align} 

Comparing the RHS of Eqs.~(\ref{relation_1_sup}) and~(\ref{relation_2_sup}) we obtain a quadratic equation of the roughness exponent $\chi$ as a function of dimension $d$ and dimensionless ratio $\gamma$ which is
\begin{align}
A\chi^2+B\chi+C=0,
\end{align}
where,
\begin{align*}
&A=36\gamma^2+(6d+30)\gamma+(3d+6),\\&B=(36d-54)\gamma^2+(8d^2+16d-42)\gamma+\biggl(\frac{d^2}{2}-4-d\biggl),\\&C=(d-2)\times\gamma\times\bigl[\gamma(8d-8)+(2d^2+4d-6)\bigl],
\end{align*}   
as given in the main text.

\end{document}